\documentclass{article}
\def\const {{\hbox{const}}}
\def\O {{\cal O}}
\newtheorem{Def}{Definition}

\title
{ Singular solution of the Liouville equation under
perturbation}
\author
{ L.A. Kalyakin
\thanks
{This research has been supported by the Russian
Foundation of the Fundamental Research under Grants
99-01-00139, 96-15-96241} }
\date{\it Institute of Mathematics RAS, Ufa, Russia}

\begin{document}
\maketitle

The Cauchy problem for the Liouville equation with a
small perturbation
$$
\partial_t^2u   -\partial_x^2u
+8\exp u =\varepsilon{\bf F}[u];
\quad 0< \varepsilon \ll 1,
\eqno (0.1)
$$
$$
u |_{t=0}=\psi_0(x;\varepsilon ), \
\partial_tu|_{t=0}=\psi_1(x;\varepsilon ),\quad x\in R
\eqno (0.2)
$$
is considered. We are interesting for asymptotics of the
perturbed solution $u(x,t;\varepsilon)$\ as
$\varepsilon\to 0$.

Perturbation theory for integrable equations remains a
very attractive task. As a rule a perturbation of smooth
solutions such as a single soliton were usually
considered. We intend here to discuss a perturbation of a
singular solution under assumption that the perturbed
solution has singularities as well. A simple well known
instance of this kind is a chock wave under weak
perturbation as given by the Hopf equation with a small
perturbation term
$
u_t+uu_x=\varepsilon f(u),\ 0< \varepsilon \ll 1.
$
In this paper we consider a more complicated problem
namely a perturbation of the singular solution of
Liouville equation. We deal with the singularities
studied by Pogrebkov and Polivanov, [1].

\bigskip
{\bf 1. Unperturbed equation} (as $\varepsilon=0$)
$
\partial_t^2u   -\partial_x^2u
+8\exp u =0$ was solved by Liouville [2] and a formula
of the general solution is well known
$$
u(x,t)=\ln {{r_+^{\prime}(s^+)r_-^{\prime}(s^-)}\over
{r^2(s^+,s^-)}},\quad  r=r_+(s^+)+r_-(s^-), \
s^{\pm}=x\pm t.
\eqno (1.1)
$$
Initial equations (0.2) give two ODE's for the $r_\pm$
which may be linearized.

The singularities of the solution occur due to zero of
the denominate $ r(x+t,x-t)\equiv r_+(x+t)+r_-(x-t)=0$
under condition $\ r_\pm^\prime >0.$ We consider just
this case. The singular solution is unique under
matching condition imposed on the singular lines
$\Gamma$, [3]. That is continuity (zero jump across
the singular lines) of two expressions
$$
[1/2(u _t^2+u _x^2)+2\exp u
-2u _{xx}]_{\Gamma}=0,\quad
 [-u _t\phi _x+ 2u _{tx}]_{\Gamma}=0.
\eqno (1.2)
$$

\bigskip
{\bf 2. Perturbed problem ($\varepsilon\neq 0$)}

As regards the perturbation operator $F[u]$ we assume
that one has no higher order derivatives. We desire
the singularities do not disperse so that singular
lines only deform slowly under perturbation. Note that
an existence theorem for the perturbed problem is not
proved up to now. We only give a formal asymptotic
solution.

The main goals are both an asymptotic approximation of
the singular lines and an asymptotic approximation of
the solution everywhere out of narrow neighborhood of
the singular lines.

\begin{Def}
The formal asymptotic solution (FAS) of order $N$ is a
function $U_N(x,t;\varepsilon )$ satisfied to both the
equation (0.1) and the initial condition (0.2) up to
order $\O(\varepsilon^N)$ everywhere in \{$x\in R, \
0\leq t\leq T=\const$\} out of narrow strips of order
$\O(\varepsilon^N)$. In the strips there are smooth
lines on which the matching conditions (1.2) are
satisfied.
\end{Def}

{\bf Remark.} A direct asymptotic expansion as given by
$
u\approx\sum_n\varepsilon^n\stackrel{n}{u}(x,t),
\quad \varepsilon\to 0
$
does not provide an approach to both the singular
lines and the solution inside of stripes (near limit
singular lines) whose width has the order of
$\O(\varepsilon)$. This assertion can be verified on a
simple example when $F\equiv 0$ and exact solution is
taken in the explicit form (1.1) with the smooth
functions $r_\pm(s^\pm;\varepsilon)$ depending on the
parameter $\varepsilon$. Asymptotic expansion of such
solution as $\varepsilon\to 0$ has coefficients with
the increasing order singularities $\O(r_0^{-n})$ at
the limit singular lines $r_0^{-n}\equiv
r_+(s^+;0)+r_-(s^-;0)=0$.

{\bf Anzatz.} A FAS (for any N) is taken as a finite
peace of the asymptotic series
$$
u(x,t;\varepsilon  )
\approx\sum_{n=0}^\infty\varepsilon^n
\stackrel{n}{u}(x,t;\varepsilon ),\quad \varepsilon\to 0.
\eqno (2.1)
$$
The leading order term is here taken as an exact solution
of the unperturbed equation
$$
\stackrel{0}{u}=
\ln {{r_+^{\prime}(s^+;\varepsilon  )
r_-^{\prime}(s^-;\varepsilon  )}\over
{r^2(s^+,s^-;\varepsilon  )}},\quad (r=r_++r_-).
\eqno (2.2)
$$
We permit dependence on a small parameter in the
functions $r_\pm(s_\pm;\varepsilon)$. For these
functions an asymptotics is constructed in the form
$$
r_\pm (s^\pm;\varepsilon  )
\approx\sum_{n=0}^\infty\varepsilon^n
\stackrel{n}{r}_\pm (s^\pm ),\quad \varepsilon\to 0.
\eqno (2.3)
$$
The formulas (2.1)--(2.3) are usually named Bogolubov
-- Krylov anzatz.  This approach provides an asymptotic approximation
of the singular lines up to any order from the
equation
$
r_+ (s^\pm;\varepsilon  )+r_- (s^\pm;\varepsilon  )=0.
$
Thus the matter is reduced to define the coefficients
$\stackrel{n}{u},\stackrel{n}{r_\pm}$.

\bigskip
{\bf 3. Linearized problem for the correction}

Corrections $\stackrel{n}{u},\ (n \geq 1)$ are determined
from the linear equations
$$
\partial_t^2\stackrel{n}{u}
-\partial_x^2\stackrel{n}{u}+
8{{r_+^{\prime}r_-^{\prime}}\over
{r^2}}\stackrel{n}{u}=\stackrel{n}{f}(x,t;\varepsilon  ),
\quad (x,t)\in R^2
$$
with the corresponding initial data. The right sides are
here calculated from previous steps, for example
$
\stackrel{1}{f}={\bf F}[\stackrel{0}{u}],\quad
\stackrel{2}{f}=\delta
{\bf F}[\stackrel{0}{u}]\stackrel{1}{u}-
4{{r_+^{\prime}r_-^{\prime}}}(\stackrel{1}{u})^2/ {r^2}.
$

General solution of the homogeneous linear equation is
given by the formula
$
u(x,t;\varepsilon )={{j_+^{\prime}}/{r_+^{\prime}}}+
{{j_-^{\prime}}/{r_-^{\prime}}}- 2{{(j_++j_-)}/{r}}.
$
Here $j_\pm=\stackrel{n}{j}_\pm(s^\pm;\varepsilon )$
are arbitrary functions. In context of the Cauchy
problem they are determined by the initial functions
and an expression similar to the D'Alembert formula
takes place. Take into account the singularities of
order $\O(1/r)$.

To solve the nonhomogeneous linear equation the similar
formula may be used with the functions  $j_\pm(x,t)$,
depending on two variables, which are defined from ODE's
in the explicit form. So a solution of the linear
equation is determined by the integral
$$
u(x,t)=\int_{s^-}^{s^+}\int_{s^-}^{\sigma^+}
K({s^+},{s^-},\sigma^+,\sigma^-)f(\sigma^+,\sigma^-)
\,d\sigma^- \,d\sigma^+
$$
taken over the characteristic triangle. The kernel $K$
is here expressed by
$$
K({s^+},{s^-},\sigma^+,\sigma^-)=
{1\over{2r(s_+,s_-)r(\sigma_+,\sigma_-)}}
\Big\{
r_+(s^+)r_-(s^-) + r_+(\sigma^+)r_-(\sigma^-)
$$
$$
 +{1\over 2}\Big[r_+(s^+)-r_-(s^-)\Big]
\Big[r_+(\sigma^+)-r_-(\sigma^-)\Big]\Big\}.
$$
Let us denote by $\stackrel{n}{u}_1(x,t;\varepsilon)$
a part of the correction corresponding to a solution
of the nonhomogeneous linear equation. One depends on
both the $r_\pm (s^\pm ;\varepsilon )$ and the
$\stackrel{m}{j}_\pm (s^\pm;\varepsilon ),\ (0<m< n)$
and one is determined by means of the quite well
specific operator $\stackrel{n}{u}_1=\stackrel{n}{\bf
U} [r_\pm,\stackrel{1}{j_\pm}
,...,\stackrel{n-1}{j_\pm}]$. It is convenient to
identify the singular part in these representation
$$
\stackrel{n}{\bf U}=\stackrel{n}{\bf V} [r_\pm,\stackrel{1}{j_\pm}
,...,\stackrel{n-1}{j_\pm} ]- {(2/r)}\stackrel{n}{\bf
W} [r_\pm,\stackrel{1}{j_\pm}
,...,\stackrel{n-1}{j_\pm} ].
\eqno (3.1)
$$

Singularities play here a role of secular terms. If we
take a FAS in the form of the direct expansion, then
we find that the singularities became stronger on each
step. One can guess that this effect is due to the
poor approximation of the perturbed singular lines.
Hence the singularities must to be eliminated from the
corrections. Just this elimination gives us a good
approximation of the perturbed singular
lines\footnote{ One can think these ideas are suitable
for another problems with singularities under
perturbations.}. Of course a representation (3.1) with
smooth functions $\stackrel{n}{\bf V},\stackrel{n}{\bf
W}$ is possible just only under some restrictions on
the perturbation operator $F[u]$. In fact, the more
general condition on the $F[u]$ is a representation
(3.1) with smooth functions $\stackrel{n}{\bf
V},\stackrel{n}{\bf W}$ on each step $n$.

\bigskip
{\bf 4. Identification of corrections}

The system of linear equations are  based on the
leading term whose parameters  $r_\pm $ are determined
by initial data from equation:
$$
\ln {{r_+^{\prime}r_-^{\prime}}\over{r^2}}
+\sum_{n=1}^\infty\varepsilon^n
\Big[{{\stackrel{n}{j_+}^\prime}/{r_+^\prime}}
+{{\stackrel{n}{j_-}^\prime}/{r_-^{\prime}}} -
{2\over{r}}
\Big(\stackrel{n}{j_+}+\stackrel{n}{j_-}\Big)
+\stackrel{n}{u_1}\Big]=\psi_0(x;\varepsilon ),
\eqno (4.1)
$$
$$
{{r_+^{\prime\prime}}\over{r_+^{\prime}}}-
{{r_-^{\prime\prime}}\over{r_-^{\prime}}}
-2{{r_+^{\prime}-r_-^{\prime}}\over{r}}+
\sum_{n=1}^\infty\varepsilon^n
\Big[({{\stackrel{n}{j_+}^\prime}/{r_+^\prime}})^\prime
-({{\stackrel{n}{j_-}^\prime}/{r_-^{\prime}}})^\prime-
{2\over{r}}
\Big(\stackrel{n}{j_+}^\prime-\stackrel{n}{j_-}^\prime\Big)+
$$
$$
+{2\over{r^2}}\Big(\stackrel{n}{j_+}+\stackrel{n}{j_-}\Big)
({r_+}^\prime-{r_-}^\prime)
+\partial_t\stackrel{n}{u_1}\Big]=\psi_1(x;\varepsilon ).
\eqno (4.2)
$$
 Note that there are unknown functions
 $\stackrel{n}{j}_\pm$ apart from $r_\pm$.

A naive approach to these equations with a small
parameter is as follows. Functions $r_\pm$ are identified
with the leading term $r_\pm=\stackrel{0}{r_\pm}(s^\pm )$
which is defined from the nonlinear equations as
$\varepsilon
=0$. After that all functions $\stackrel{n}{j_\pm}(s^\pm )$
are determined from the recurrent system of linear
equations. In this way a direct asymptotic expansion
is just obtained, whose coefficients have the
singularities of increasing order $\O(r_0^{-n})$ at
the limit singular lines $r_0\equiv
r_+(s^+;0)+r_-(s^-;0)=0$. Deformation of the singular
lines is not determined in this way.

In our approach the functions $r_\pm$ are taken in the
form of asymptotic series (2.3). Additional ambiguities
in the coefficients $\stackrel{n}{r_\pm}(s^\pm ),\ n\geq
1$ are used to solve the (4.1),(4.2) under additional
requirements. We desire to eliminate the singularities of
order $\O(r^{-1})$ from the corrections $\stackrel{n}{u}$
on each step. Elimination of these terms give rise to the
algebraic equations
$$
\stackrel{n}{j_+}(s^+;\varepsilon )
+\stackrel{n}{j_-}(s^-;\varepsilon )+\stackrel{n} {\bf
W} [r_\pm,\stackrel{1}{j_\pm}
,...,\stackrel{n-1}{j_\pm} ]=0\quad  {\rm as} \quad
r(s^+,s^-;\varepsilon )=0.
\eqno (4.3)
$$
Moreover an additional condition is used
$$
\stackrel{n}{r_+}(x)+\stackrel{n}{r_-}(x)=0, \quad
\forall x\in R, \quad (n\geq 1).
\eqno (4.4)
$$
So we obtain the systems of equations (4.1)--(4.4) for
the four functions
$\stackrel{n}{r_\pm},\stackrel{n}{j_\pm}$ on each step
$n=1,2,...$. However these equations contain a small
parameter throw $r_\pm (x;\varepsilon)$. Hence an
asymptotic expansion have to be constructed for the
functions $\stackrel{n}{j_\pm}$ like (2.3)
$
\stackrel{n}{j_\pm }(x;\varepsilon )
\approx\sum_{m}\varepsilon^m
\stackrel{n,m}{j_\pm }(x).
$
Coefficients can be here obtained from the recurrent
system of algebraic equations
$$
\stackrel{n,m}{j_+}(s^+)+\stackrel{n,m}{j_-}(s^-)
=\stackrel{n,m}{W}, \quad (n\geq 1,\ 0<m<n)\quad {\rm as} \ \stackrel{0}{r}(s^+,s^-)=0.
\eqno (4.5)
$$

So all functions
$\stackrel{n}{r_\pm},\stackrel{n,0}{j_\pm}(x),\ (n\geq
1)$\ and\ $\stackrel{n,m}{j_\pm}\ (1\leq m<n$)\  are
defined step--by--step from the algebraic and
differential equations. The main objects are differential
equations, which can be represented in the form
$$
{{\stackrel{n}{y_+}^{\prime}}/{\stackrel{0}{r_+}^{\prime}(x)}}+
{{\stackrel{n}{y_-}^{\prime}}/{\stackrel{0}{r_-}^{\prime}(x)}}-
2{{\stackrel{n}{y_+}+\stackrel{n}{y_-}}\over{r(x,x)}}
=\stackrel{n}{\Psi}_0(x),
$$
$$
\Big({{\stackrel{n}{y_+}^{\prime}}/{\stackrel{0}{r_+}^{\prime}(x)}}
-{{\stackrel{n}{y_-}^{\prime}}/{\stackrel{0}{r_-}^{\prime}(x)}}\Big)^{\prime}-
2{{\stackrel{n}{y_+}^{\prime}-\stackrel{n}{y_-}^{\prime}}\over{r(x,x)}}
+2{{(\stackrel{n}{y_+}+\stackrel{n}{y_-})
(\stackrel{0}{r_+}^{\prime}(x)-\stackrel{0}{r_-}^{\prime}(x))}\over{r^2(x,x)}}=
\stackrel{n}{\Psi}_1(x)
$$
for the combinations
$
\stackrel{n}{y}_\pm (x)=
\stackrel{n}{r}_\pm (x)+\stackrel{n,0}{j}_\pm (x).
$
If we take into account
${r}(x,x)=\stackrel{0}{r_+}(x)+\stackrel{0}{r_-}(x)$,
then a solution is obtained in the explicit form
$$
y_{\pm}={1\over 2}\int(\Psi_0\pm
g)\stackrel{0}{r_{\pm}}^{\prime}\,dx +{1\over 2}\int
\stackrel{0}{r_{\pm}}^{\prime}
\int\Big({\Psi_0{{{r}} ^{\prime}}/{r}}+
{{(\stackrel{0}{r}_+
-\stackrel{0}{r}_-)^{\prime}} g/
{{r}}}\Big)\,dx\,dx,
$$
$$
g(x)={r}(x,x)\int\Big({{\Psi_1(x)}
/{{r}(x,x)}}
+{{\Psi_0(x)\Big(r_+(x)-r_-(x)
\Big)^{\prime} }/{{r}^2(x,x)} } \Big)\, dx.
$$
After that the algebraic equations (4.4)--(4.5) are
solved. There are some arbitrariness in the solution
$\stackrel{n}r_\pm,\stackrel{n,m}j_\pm$, which however do
not effect the obtained approximation for the
$u(x,t;\varepsilon)$.

{\bf Conclusion.} The main result of the paper is the
given above manner of determination of the FAS
(2.1)--(2.3) in the occurence of singularities.

\end{document}